\documentclass{article}
\usepackage{graphicx} 
\usepackage{amsthm, amsmath, amsfonts,mathtools, fullpage, graphicx, wrapfig, tikz, caption, subcaption}
\usepackage{natbib}
\usepackage{amssymb}
\usepackage{verbatim}
\usepackage{relsize}
\usepackage{listings}
\usepackage{xcolor}
\usepackage{hyperref}
\usepackage{graphicx}
\usepackage{multirow}
\usepackage[linesnumbered,ruled,vlined]{algorithm2e}

\newtheorem{assumption}{Assumption}
\usepackage{booktabs}
\usepackage{tabularx}

\usepackage{setspace}

 \usepackage[vmargin=1.18in,hmargin=1.1in]{geometry}

\title{A Note on The Rationale Behind Using Parental Longevity as a Proxy in Mendelian Randomization Studies}
\author{Zach Shahn$^{1,2}$, Rehana Rasul$^{1}$, and Mary Schooling$^{1}$\\
$^1$:CUNY Graduate School of Public Health and Health Policy, New York, NY, United States\\
$^2$:Institute for Implementation Science in Population Health, New York, NY, United States}

\begin{document}

\maketitle

In many cohorts (such as the UK Biobank \citep{sudlow2015uk}) on which Mendelian Randomization (MR) studies are routinely performed, data on participants' longevity is inadequate as the majority of participants are still living. To nevertheless estimate effects on longevity, it is increasingly common for researchers to substitute participants' `parental attained age', i.e. parental lifespan or current age (which is routinely collected in UK Biobank), as a proxy outcome \citep{pilling2017human, timmers2019genomics,ye2023mendelian,ng2023effect}. The common approach to performing this clever trick appears to be based on a solid understanding of its underlying assumptions. However, we have not seen these assumptions (or the causal effects whose identification they enable) clearly stated anywhere in the literature. In this note, we fill that gap, with our main conclusions summarized in Table \ref{tab:placeholder}. 



Suppose that we have available a sample of independent and identically distributed realizations of the random variables $(G,A,Y_P)$, where $G$ denotes presence of a genetic variant or SNP in the population of interest, $A$ denotes the exposure of interest in the population of interest, and $Y_P$ denotes an outcome of interest (usually longevity, but others are also possible) measured in a parent. Relevant variables that are unobserved are: $G_P$, the variant in the parents; $A_P$, the exposure in the parents; and $Y$, the outcome in the population of interest. 

\begin{figure}[h]
\centering
\includegraphics[scale=.75]{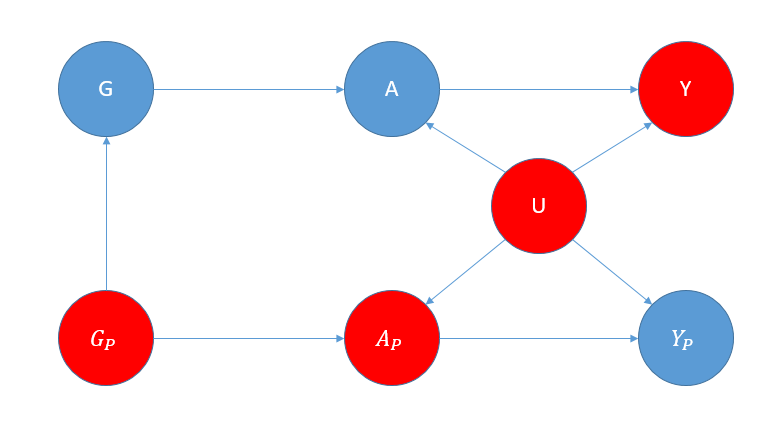}  
\caption{Directed Acyclic Graph depicting causal assumptions underlying use of parental proxies in Mendelian Randomization studies. Observed variables are in blue and unobserved variables are in red. The DAG, along with the faithfulness assumption, implies that $G$ is a valid instrument for the $A$-$Y$ relationship in the population of interest, $G_P$ is a valid instrument for the $A_P$-$Y_P$ relationship in the parents, and $G$ is a valid (non-causal) instrument for the $A_P$-$Y_P$ relationship.}
\label{dag}
\end{figure}

We assume that the causal Directed Acyclic Graph (DAG) in Figure \ref{dag} describes the relationships among these variables. The DAG (plus the faithfulness assumption \citep{spirtes2000causation}) has the following implications:
\begin{assumption}\label{g_y_iv}
    $G$ is a valid instrumental variable for the effect of $A$ on $Y$.
\end{assumption}
\begin{assumption}\label{gp_yp_iv}
    $G_P$ is a valid instrumental variable for the effect of $A_P$ on $Y_P$.
\end{assumption}
\begin{assumption}\label{g_yp_iv}
    $G$ is a valid (non-causal) instrumental variable for the effect of $A_P$ on $Y_P$.
\end{assumption} 
\noindent When we say that an exposure is a `valid instrumental variable' for an outcome, we mean that the three instrumental variable conditions---(IV.1) relevance, (IV.2) exclusion restriction, and (IV.3) unconfoundedness---are satisfied. Relevance states that the exposure and instrument are associated. The DAG in Figure \ref{dag} and faithfulness imply the relevance component of Assumptions \ref{g_y_iv}-\ref{g_yp_iv} because the instruments and exposures are not d-separated. The exclusion restriction states that the instrument has no direct effect on the outcome not through the exposure.  Figure \ref{dag} implies the exclusion restriction component of Assumptions \ref{g_y_iv}-\ref{g_yp_iv} via the absence of any directed paths from instrument to outcome that do not pass through exposure. Finally, the unconfoundedness condition states that the instrument and outcome are not associated (e.g. through common causes or collider paths opened due to selection) except through the exposure. Figure \ref{dag} implies the unconfoundedness component of Assumptions \ref{g_y_iv}-\ref{g_yp_iv} via the absence of any open backdoor paths between instruments and outcomes. We presume basic familiarity with instrumental variables for the purposes of this note and refer readers to Chapter 16 of \citet{hernan2010causal} for a thorough review.

We note that Assumptions \ref{g_y_iv} and \ref{gp_yp_iv}---though they concern the same variants, exposures, and outcomes in different populations---do not necessarily imply one another. Consider the following contrived example. Suppose that the SNP represented by $G$ and $G_P$ causes two life threatening conditions---one represented by $A/A_P$ and another represented by $B/B_P$. (Here, as before, $B$ is the condition in the population of interest, and $B_P$ is the condition in the parents.) Further suppose that an effective childhood vaccine for condition $B$ was discovered shortly before the population of interest were born. Then, in the parents' era, the pathway $G_P\rightarrow B_P\rightarrow Y_P$ would have violated the exclusion restriction and thus Assumption \ref{gp_yp_iv}. However, in the population of interest, this pathway would be absent due to the vaccine and the exclusion restriction with respect to $A$ and Assumption \ref{g_y_iv} could still hold.

Assumptions \ref{gp_yp_iv} and \ref{g_yp_iv}, on the other hand, are essentially equivalent. Assumption \ref{g_yp_iv} is an immediate consequence of Assumption \ref{gp_yp_iv} and basic biology. Furthermore, Assumption \ref{g_yp_iv} would not hold if Assumption \ref{gp_yp_iv} did not hold. This is because: if $G_P$ is not associated with $A_P$ (violating the relevance condition of Assumption \ref{gp_yp_iv}), then $G$ would also not be associated with $A_P$ (violating the relevance condition of Assumption \ref{g_yp_iv}); and if there is any open path between $G_P$ and $Y_P$ not through $A_P$ (violating the exclusion or independence conditions of Assumption \ref{gp_yp_iv}), then that path with $G\leftarrow G_P$ appended to it would constitute an open path between $G$ and $Y_P$ not through $A_P$ (violating the independence condition of Assumption \ref{g_yp_iv}).   

\textbf{\textit{Under Assumption \ref{g_yp_iv}, an association between $G$ and $Y_P$ implies that there is a causal relationship between $A_P$ and $Y_P$}}. That is, it implies that the exposure of interest caused the outcome of interest \textit{in the parents}. This does not, however, necessarily imply that it causes the outcome in the population of interest. Similarly, while the absence of an association between $G$ and $Y_P$ implies that the exposure did not cause the outcome in the parents, the exposure still might cause the outcome in the population of interest.  As another contrived example of how effects can change over time, suppose that the population of interest experienced a famine while their parents did not. High adiposity could then be protective in the population of interest but not the parents. Thus, stability of effects over time is an extra assumption required to interpret an association between $G$ and $Y_P$ as evidence of an effect of $A$ on $Y$. We would expect most biological effects to persist over generations, though, barring exceptional circumstances.

Many researchers do not stop at examining the association between $G$ and $Y_P$. Instead, they seek to estimate an effect magnitude. If they were able to observe $Y$, one way to go about this would be to estimate the Wald functional
\begin{equation}\label{Wald}
    \frac{E[Y|G=1]-E[Y|G=0])}{E[A|G=1]-E[A|G=0]}.
\end{equation}
For identification of effect magnitude, an additional condition is required beyond the IV conditions (IV.1)-(IV.3). Under Assumption \ref{g_y_iv} plus an effect homogeneity condition, (\ref{Wald}) identifies the average effect of treatment on the treated (ETT). Under Assumption \ref{g_y_iv} plus a monotonicity condition, (\ref{Wald}) identifies the Local Average Treatment Effect. (See Chapter 16 of \citet{hernan2010causal} for a full discussion of IV estimands and assumptions.)  As $Y$ is not observed, researchers instead compute
\begin{equation}\label{Wald1}
    \frac{f(E[Y_P|G=1]-E[Y_P|G=0])}{E[A|G=1]-E[A|G=0]},
\end{equation}
where $f(\cdot)$ is a function that converts the association between $G$ and $Y_P$ to an implied (based on knowledge of genetic mechanisms) association between $G_P$ and $Y_P$. Given that a child inherits one of two parental alleles at random, the association involving the child's genotype ($G$) would be, on average, half the magnitude of the association involving the parent's own genotype ($G_P$). Under Mendelian inheritance, then, $f(x) \approx 2x$. 

\textbf{\textit{Under Assumption \ref{g_y_iv}, effect homogeneity or monotonicity in the population of interest, and the assumption that the unobserved $G$-$Y$ association is equal to the $G_P$-$Y_P$ association, i.e.}}
\begin{equation}\label{outcome_stab}
  E[Y|G=1]-E[Y|G=0]=E[Y_P|G_P=1]-E[Y_P|G_P=0], 
\end{equation}
\textbf{\textit{(\ref{Wald1}) identifies the effect (LATE or ETT, depending on the fourth condition) in the population of interest}}. This follows simply because (\ref{outcome_stab}) implies that (\ref{Wald1})=(\ref{Wald}), and the remaining assumptions imply that (\ref{Wald}) is equal to the desired effect.

\textit{\textbf{Under Assumptions \ref{gp_yp_iv} and \ref{g_yp_iv}, effect homogeneity or monotonicity in the parental population, and the alternative assumption that the SNP-exposure association is constant over time, i.e.}}
\begin{equation}\label{exp_stab}
E[A|G=1]-E[A|G=0]=E[A_P|G_P=1]-E[A_P|G_P=0],    
\end{equation}
\textbf{\textit{(\ref{Wald1}) identifies the ETT or LATE in the parents.}} The reasoning is similar to before. (\ref{exp_stab}) implies that (\ref{Wald1}) is equal to 
\begin{equation*}
    \frac{E[Y_P|G_P=1]-E[Y_P|G_P=0]}{E[A_P|G_P=1]-E[A_P|G_P=0]},
\end{equation*}
which, under Assumption \ref{gp_yp_iv} and an effect homogeneity or monotonicy condition \textit{in the parents} is equal to the ETT or LATE in the parents.

\begin{table}[ht]
    \centering
    \renewcommand{\arraystretch}{1.3} 
    \begin{tabularx}{\textwidth}{lX}
        \toprule
        \textbf{Goal} & \textbf{Requires} \\
        \midrule
        Effect direction in parents &
            Assumption~\ref{g_yp_iv} \\
        Effect direction in participants &
            Assumption~\ref{g_yp_iv}, stability of directional effects across generations \\
        Effect magnitude in parents &
            Assumptions \ref{gp_yp_iv} and ~\ref{g_yp_iv}, homogeneity/monotonicity in parents, and stable gene--exposure association across generations (\ref{exp_stab}) \\
        Effect magnitude in participants &
            Assumption~\ref{g_y_iv}, homogeneity/monotonicity in participants, and stable gene--outcome association across generations (\ref{outcome_stab}) \\
        \bottomrule
    \end{tabularx}
    \caption{Summary of alternative estimands and required assumptions in MR analyses with parental outcomes as proxies}
    \label{tab:placeholder}
\end{table}

Table 1 provides a summary of the various assumptions required for MR using parental outcome proxies to target various quantities of interest. We would argue that it is unlikely that stability of the SNP-outcome association (\ref{outcome_stab}) would hold without stability of the SNP-exposure association (\ref{exp_stab}), as the outcome association flows through the exposure association. Therefore, we find (\ref{exp_stab}), which only allows identification of the effect in the parents, a somewhat weaker assumption. However, the stable SNP-exposure association assumption (\ref{exp_stab}) is still unlikely to hold given changes in diet, culture, medicine, etc. over time. One might speculate, though, that the gene-exposure association for a highly heritable biomarker less influenced by lifestyle (e.g., lipoprotein(a)) may be more stable across generations than the association for exposures like BMI or educational attainment, which are heavily modified by environmental and social changes. Even if the magnitude of an effect on longevity in the parents were accurately estimated, we would expect the magnitude to be different in the children. For example, exposures causing heart disease had larger effects on longevity in past generations as treatments have improved over time. Tests for presence and directionality of an effect in the parents via the association between children's genes and parents' outcomes should be more robust, with the caveat that these effects may also not have persisted over time. Even if an analysis successfully identifies a causal effect in the parents, this is ultimately a historical causal claim as the primary output of this analysis is a statement about biology and environment in a previous generation. Using that result to inform policy or clinical understanding for the current generation requires an additional step that relies on questionable stability assumptions.

\bibliography{bibliography}
\end{document}